\documentclass[12pt,preprint]{aastex}

\usepackage{graphicx}
\usepackage{rotating,subfigure}

\shorttitle{IRAM-PdBI Observations of Binary Protostars I}
\shortauthors{Chen et al.}

\begin{document}

%%%%%%%%%%%%%%%%%%%%%%%%%%%%%%%%%%%%%%%%%%%%%%%%%%%%%%%%%%%%%%%%%%%%%%%%%%%%%%%%%%%%%%%%%%%%%%
\title{IRAM-PdBI Observations of Binary Protostars I: The Hierarchical System SVS\,13 in NGC\,1333}
%%%%%%%%%%%%%%%%%%%%%%%%%%%%%%%%%%%%%%%%%%%%%%%%%%%%%%%%%%%%%%%%%%%%%%%%%%%%%%%%%%%%%%%%%%%%%%

\author{Xuepeng Chen, Ralf Launhardt, and Thomas Henning}
\affil{Max-Planck-Institut f\"{u}r Astronomie, K\"{o}nigstuhl 17, D-69117 Heidelberg, Germany}
\email{chen@mpia.de}

\begin{abstract}

We present millimeter interferometric observations of the young
stellar object SVS\,13 in NCG\,1333 in the N$_{2}$H$^{+}$\,(1--0)
line and at 1.4 and 3\,mm dust continuum, using the IRAM Plateau 
de Bure interferometer. The results are complemented by infrared 
data from the {\it Spitzer Space Telescope}. The millimeter dust 
continuum images resolve four sources (A, B, C, and VLA\,3) in 
SVS\,13. With the dust continuum images, we derive gas masses of 
0.2\,$-$\,1.1\,$M_\odot$ for the sources. N$_{2}$H$^{+}$\,(1--0) 
line emission is detected and spatially associated with the dust 
continuum sources B and VLA\,3. The observed mean line width is 
$\sim$\,0.48\,km\,s$^{-1}$ and the estimated virial mass is 
$\sim$\,0.7\,$M_\odot$. By simultaneously fitting the seven 
hyperfine line components of N$_{2}$H$^{+}$, we derive the velocity 
field and find a symmetric velocity gradient
of $\sim$\,28\,km\,s$^{-1}$\,pc$^{-1}$ across sources B and
VLA\,3, which could be explained by core rotation. The velocity field 
suggests that sources B and VLA\,3 are 
forming a physically bound protobinary system embedded in a common
N$_{2}$H$^{+}$ core. $Spitzer$ images show mid-infrared emission
from sources A and C, which is spatially associated with the mm
dust continuum emission. No infrared emission is detected from
source B, implying that the source is deeply embedded. 
Based on the morphologies and velocity structure, we propose a
hierarchical fragmentation picture for SVS\,13 where the three
sources (A, B, and C) were formed by initial fragmentation of a
filamentary prestellar core, while the protobinary system (sources
B and VLA\,3) was formed by rotational fragmentation of a single
collapsing sub-core.

\end{abstract}


\newpage
%%%%%%%%%%%%%%%%%%%%%%%%%%%%%%%%%%%%%%%%%%%%%%%%%%%%%%%%%%%%%%%%%%%%%%%%%%%%%%%%%%%%%%%
\section{INTRODUCTION}
%%%%%%%%%%%%%%%%%%%%%%%%%%%%%%%%%%%%%%%%%%%%%%%%%%%%%%%%%%%%%%%%%%%%%%%%%%%%%%%%%%%%%%%

Although both observations and theoretical simulations support the
hypothesis that the fragmentation of collapsing protostellar cores
is the main mechanism for the formation of binary/multiple stellar
systems, many key questions concerning this fragmentation process,
e.g., the exact {\it when}, {\it where}, {\it why}, and {\it
how}, are still under debate (see reviews by Bodenheimer et al.
2000, Tohline 2002, and Goodwin et al. 2007). To answer these
questions, direct observations of the earliest, embedded phase of
binary star formation are needed to study in detail their
kinematics. This was unfortunately long hampered by the low angular
resolution of millimeter (mm) telescopes. However, the recent
availability of large millimeter interferometers has enabled us to
directly observe the formation phase of binary stars, although the
number of known and well-studied systems is still very small (see
Launhardt 2004).

To search for binary protostars and to derive their kinematic
properties, we have started a systematic program to observe, at
high angular resolution, a number of isolated low-mass (pre-)
protostellar molecular cloud cores. The initial survey was
conducted at the Owens Valley Radio Observatory (OVRO)
millimeter-wave array (Launhardt 2004; Chen et al.
2007, hereafter Paper\,I; Launhardt et al. in prep.), and is now
continued with the Australia Telescope Compact Array (ATCA; Chen
et al. 2008a, hereafter Paper\,II), the Submillimeter Array (SMA;
Chen et al. 2008b), and the IRAM Plateau de Bure Interferometer
(PdBI) array (this work and Chen et al. in prep.).

%\subsection{SVS 13}
SVS\,13 is a young stellar object (YSO) located in the NGC\,1333
star-forming region at a distance of 350\,pc\footnote{Although
recent VLBA observations suggest a distance of 220\,pc (see Hirota
et al. 2007), we use here 350\,pc for consistency with earlier
papers.} (Herbig \& Jones 1983). It was discovered as a
near-infrared (NIR) source by Strom et al. (1976). At least three
mm continuum sources were detected around SVS\,13 (Chini et al.
1997; Bachiller et al. 1998, hereafter B98) and named A, B, and C,
respectively (Looney et al. 2000, hereafter LMW2000). Source A is
coincident with the infrared/optical source SVS\,13, source B is
located $\sim$\,15$''$ southwest of A, and source C is further to the
southwest (B98; LMW2000). With BIMA observations,
LMW2000 detected another mm source located $\sim$\,6$''$ southwest
of source A, which is coincident with the radio source VLA\,3 in
the VLA survey of this region (Rodr\'{i}guez et al. 1997; 1999).
More recently, Anglada et al. (2000; 2004) revealed that source A
is actually a binary system with an angular separation of
0\farcs3. Located to the southeast of SVS\,13 is the well-known
chain of Herbig-Haro (HH) objects 7$-$11 (Herbig 1974; Strom et al.
1974; Khanzadyan et al. 2003 and references therein). Although
several possible driving candidates have been proposed for this
large-scale HH chain (e.g., Rodr\'{i}guez et al. 1997),
high-resolution CO\,(2--1) observations (Bachiller et al. 2000,
hereafter B2000), as well as an analysis of the spectral energy
distributions (LMW2000), clearly favor SVS13\,A as the driving source.
In this paper, we present our IRAM-PdBI observations in the
N$_{2}$H$^{+}$\,(1--0) line and at the 1.4 and 3\,mm dust continuum
toward SVS\,13, together with complementary infrared data from the
{\it Spitzer Space Telescope} (hereafter $Spitzer$).
In Section\,2 we describe the observations and data reduction;
Observational results are presented in Section\,3 and discussed in
Section\,4; The main conclusions of this study are summarized in
Section\,5.

%%%%%%%%%%%%%%%%%%%%%%%%%%%%%%%%%%%%%%%%%%%%%%%%%%%%%%%%%%%%%%%%%%%%%%%%%%%%%%%%%%%%%%%
\section{OBSERVATIONS AND DATA REDUCTION}
%%%%%%%%%%%%%%%%%%%%%%%%%%%%%%%%%%%%%%%%%%%%%%%%%%%%%%%%%%%%%%%%%%%%%%%%%%%%%%%%%%%%%%%

\subsection{IRAM PdBI Observations}

Millimeter interferometric observations of SVS\,13 were carried
out with the IRAM\footnote{IRAM is supported by INSU/CNRS
(France), MPG (Germany), and IGN (Spain).} PdBI in 2006 March (C
configuration with 6 antennas) and July (D configuration with 5
antennas). The 3\,mm and 1\,mm bands were observed simultaneously,
with baselines ranging from 16\,m to 176\,m. During the
observations, two receivers were tuned to the N$_{2}$H$^{+}$
(1--0) line at 93.1378\,GHz and the $^{13}$CO\,(2--1) line
at 220\,GHz, respectively. Bandwidths in the N$_{2}$H$^{+}$ line and
$^{13}$CO line were 20\,MHz and 40\,MHz, resulting in channel
spacings of 39\,kHz and 79\,kHz and velocity resolutions of
0.2\,km\,s$^{-1}$ and 0.1\,km\,s$^{-1}$, respectively. The
remaining windows of the correlator were combined to observe the
dust continuum emission with a total band width of 500\,MHz at
both $\lambda$\,3\,mm and $\lambda$\,1.4\,mm. System
temperatures of the 3\,mm and 1\,mm receivers were typically
110\,$-$\,200\,K and $\sim$\,300\,K, respectively. The
(naturally-weighted) synthesized half-power beam widths were
2\farcs8\,$\times$\,2\farcs5 at 93.2\,GHz and
1\farcs2\,$\times$\,1\farcs1 at 220\,GHz. The FWHM primary beams
were $\sim$\,54$''$ and 23$''$, respectively. Several nearby phase
calibrators were observed to determine the time-dependent complex
antenna gains. The correlator bandpass was calibrated with the
sources 3C273 and 1749+096, while the absolute flux density scale
was derived from 3C345. The flux calibration uncertainty was estimated
to be $\sim$\,15\%. Only N$_{2}$H$^{+}$ line data were used here
because the signal-to-noise ratio of the $^{13}$CO line data was
much lower in the 3\,mm observational conditions, although the
quality of the 1.4\,mm dust continuum data is sufficient for our
science goals. The data were calibrated and reduced using the
GILDAS\footnote{http://www.iram.fr/IRAMFR/GILDAS} software.
Observing parameters are summarized in Table~1.

\subsection{\emph{Spitzer} Observations}

Mid-infrared data were obtained from the $Spitzer$ Science
Center\footnote{http://ssc.spitzer.caltech.edu}. SVS\,13 was
observed by $Spitzer$ on 2004 September 8th with the Infrared
Array Camera (IRAC; AOR key 5793280) and September 20th with the
Multiband Imaging Photometer for $Spitzer$ (MIPS; AOR key
5789440). The source was observed as part of the c2d Legacy
program (Evans et al. 2003). The data were processed by the
$Spitzer$ Science Center using their standard pipeline (version
S14.0) to produce Post Basic Calibrated Data (P-BCD) images, which
are flux-calibrated into physical units (MJy sr$^{-1}$). Further
analysis and figures were completed with the IRAF and GILDAS
software packages.

%%%%%%%%%%%%%%%%%%%%%%%%%%%%%%%%%%%%%%%%%%%%%%%%%%%%%%%%%%%%%%%%%%%%%%%%%%%%%%%%%%
\section{RESULTS}
%%%%%%%%%%%%%%%%%%%%%%%%%%%%%%%%%%%%%%%%%%%%%%%%%%%%%%%%%%%%%%%%%%%%%%%%%%%%%%%%%%
\subsection{IRAM-PdBI Results}
%%%%%%%%%%%%%%%%%%%%%%%%%%%%%%%%%%%%%%%%%%%%%%%%%%%%%%%%%%%%%%%%%%%%%%%%%%%%%%%%%%
%%%%%%%%%%%%%%%%%%%%%%%%%%%%%%%%%%%%%%%%%%%%%%%%%%%%%%%%%%%%%%%%%%%%%%%%%%%%%%%%%%
\subsubsection{Millimeter Dust Continuum Emission}
%%%%%%%%%%%%%%%%%%%%%%%%%%%%%%%%%%%%%%%%%%%%%%%%%%%%%%%%%%%%%%%%%%%%%%%%%%%%%%%%%%

The 3\,mm dust continuum image (see Fig.\,1) shows three distinct
sources in SVS\,13. Following LMW2000, the sources are labeled A,
B, and C, respectively. Source C, the weakest one, lies outside
the field of view at 1.4\,mm. The large-scale common envelope of
the three sources, detected in the submm (Chandler \& Richer 2000)
and mm (Chini et al. 1997) single-dish maps, is resolved out by
the interferometer at 3\,mm and 1.4\,mm. From Gaussian $uv$ plane
fitting, we derive flux densities and FWHM sizes of the sources
(see Table~2). The measured angular separations are
14\farcs6\,$\pm$\,0\farcs2 between sources A and B, and
19\farcs8\,$\pm$\,0\farcs2 between B and C. A weak emission peak,
spatially coincident with the radio source VLA\,3 (Rodr\'{i}guez
et al. 1997) and the 2.7\,mm dust continuum source A2 (LMW2000),
is detected in the 3\,mm dust continuum image, but not seen in the
higher-resolution 1.4\,mm image (see Fig.\,1). Hereafter, we refer
to this weak 3\,mm continuum source as VLA\,3.

Assuming that the mm dust continuum emission is optically thin,
the total gas mass in the circumstellar envelope was calculated
with the same method described in Launhardt \& Henning (1997). In
the calculations, we adopt an interstellar hydrogen-to-dust mass
ratio of 110 (Draine \& Lee 1984) and a factor of 1.36 accounting
for helium and heavier elements. For all sources, we use mass-averaged
dust temperature and opacity of $T_{\rm d}$ = 20\,K\footnote{To
approximately correct the derived masses for the actual dust
temperature, the mass has to be scaled with
$T_{\rm d}$(adopted)/$T_{\rm d}$ for $T_{\rm d}$ $>$ 20\,K.
This also means that the derived masses are not very sensitive
to the adopted temperature in the range between 20 and 50\,K.} and
$\kappa_{\rm 1.3mm}$ = 0.8\,cm$^2$\,g$^{-1}$ [a typical value
suggested by Ossenkopf \& Henning (1994) for coagulated grains
in protostellar cores], respectively. From the 1.4\,mm dust continuum image,
the total gas masses of sources A and B are estimated to be
0.75\,$\pm$\,0.12\,$M_\odot$ and 1.05\,$\pm$\,0.16\,$M_\odot$,
respectively. The gas masses estimated from the 3\,mm dust
continuum image are consistent with those derived from the
1.4\,mm image with a dust spectral index of $\beta$ = 0.5
($\kappa_{\nu} \propto \nu^{\beta}$). We note that this index
is slightly smaller than the typical $\beta$ = 1, but agrees with
values for beta found in other highly embedded objects
(see Ossenkopf \& Henning 1994).

%%%%%%%%%%%%%%%%%%%%%%%%%%%%%%%%%%%%%%%%%%%%%%%%%%%%%%%%%%%%%%%%%%%%%%%%%%%%%%%%%%
\subsubsection{N$_2$H$^+$\,(1--0) Emission}
%%%%%%%%%%%%%%%%%%%%%%%%%%%%%%%%%%%%%%%%%%%%%%%%%%%%%%%%%%%%%%%%%%%%%%%%%%%%%%%%%%

Figure~2 shows the velocity-integrated N$_2$H$^+$ intensity image
of SVS\,13, together with outflow information from B98 and B2000.
The emission was integrated over all seven components
of the N$_2$H$^+$ line with frequency masks that completely cover
velocity gradients within the object. A N$_2$H$^+$ core, spatially
associated with the mm continuum sources B and VLA\,3, is found in
the image. The core is elongated in the northeast-southwest
direction and double-peaked: one peak is coinciding with source B
and the other is located $\sim$\,2$''$ southwest to VLA\,3. The
FWHM radius of the whole core is measured to be $\sim$\,1520\,AU. A
jet-like extension is seen at the western edge of the core, which is
$\sim$\,10$''$ in length and extending to northwest. Three smaller
clumps (at $\sim$\,3--4\,$\sigma$ level), one located close to source
C and the other two located $\sim$\,10$''$ northeast of source A, are
also seen in the image.
We note that the small clumps towards northeast of source A are also
seen in the BIMA\,+\,FCRAO N$_2$H$^+$ image of NCG\,1333 (see Walsh
et al. 2007), and may be the remnant of a large-scale N$_2$H$^+$
envelope (see $\S$\,4.4).

Figure~3 shows the N$_2$H$^+$\,(1--0) spectra at the two peak
positions. All seven hyperfine lines of N$_2$H$^+$\,(1--0)
have been detected. Using the hyperfine fitting program in
CLASS, we derive LSR velocities ($V_{\rm LSR}$), intrinsic line
widths ($\triangle$$v$; corrected for instrumental effects), total
optical depth ($\tau_{\rm tot}$), and excitation temperatures
($T_{\rm ex}$) (see Table~3).

By simultaneously fitting the seven hyperfine components with the
routine described in Paper\,I, we derive the mean velocity field
of SVS\,13 (see Fig.\,4). The outflow information (from B98) is also
shown in the map. SVS\,13 shows a well-ordered velocity field, with a
smooth gradient from southwest to northeast. A least-squares
fitting of the velocity gradient has been performed
with the routine described in Goodman et al. (1993) and provides the
following results: the mean core velocity is
$\sim$\,8.6\,km\,s$^{-1}$, the velocity gradient is
28.0\,$\pm$\,0.1\,km\,s$^{-1}$\,pc$^{-1}$, and the position angle
(P.A.) of the gradient is 51.3\,$\pm$\,0.2 degree. The details of
the velocity field are discussed in $\S$\,4.2.

Figure~5 shows the spatial distribution of the N$_{2}$H$^{+}$ line
widths in the map. We find that the line widths are roughly
constant within the core and large line widths are mainly seen in
the gap between the two emission peaks and the jet-like extension
at the core edge. The mean line width
(0.48\,$\pm$\,0.01\,km\,s$^{-1}$) is derived from Gaussian
fitting to the distribution of line widths versus solid
angle area (see Fig.\,6). The virial mass of the N$_{2}$H$^{+}$
core is then estimated to be 0.7\,$\pm$\,0.1\,$M_\odot$, using the
same method described in Paper\,II. The derived virial mass is
slightly smaller than but still comparable to the total gas mass
derived from the mm dust continuum images for source B.

%%%%%%%%%%%%%%%%%%%%%%%%%%%%%%%%%%%%%%%%%%%%%%%%%%%%%%%%%%%%%%%%%%%%%%%%%%%%
\subsection{{\it Spitzer} Results}
%%%%%%%%%%%%%%%%%%%%%%%%%%%%%%%%%%%%%%%%%%%%%%%%%%%%%%%%%%%%%%%%%%%%%%%%%%%%

Figure~7 shows the $Spitzer$ images of SVS\,13. In a wide-field
IRAC\,2 (4.5\,$\mu$m) image shown in Fig.\,7a, a number of HH
objects are seen, implying active star formation in the NGC\,1333
region. The HH objects that can be clearly distinguished are
labeled by the same numbers as in Bally et al. (1996; for a
comparison see optical image in their Figure~2). Figs.\,7b and 7c
show enlarged views of the 4.5\,$\mu$m image, overlaid with the
contours from the PdBI 3\,mm dust continuum and N$_2$H$^+$ images.
In the IRAC maps, a strong infrared source is spatially coincident
with the dust continuum source A. Located to the southeast of
source A is the well-known HH chain 7$-$11 (see Fig.\,7b),
which is associated with a high-velocity blue-shifted CO outflow (B2000).
Another collimated jet is found in the southern part of SVS\,13 with the
continuum source C being located at the apex (see Fig.\,7b), but
no near-infrared source is found at this position. Fig.\,7c shows that
the N$_2$H$^+$ emission is not spatially associated with the
infrared emission from source A, but extends roughly perpendicular
to the directions of the jets/outflows.

In the MIPS\,1 (24\,$\mu$m) image, a strong infrared source
(saturated in the image) is again found at the continuum source A,
and another weak one is found at source C. In the MIPS\,2
(70\,$\mu$m) image, extended emission are seen at both sources A
and C. It should be noted that no infrared emission is detected
from source B in any IRAC or MIPS bands, suggesting that this source is a
deeply embedded object. Flux densities of sources A and C are
measured with aperture photometry in the IRAF APPHOT package, with
the radii, background annuli, and aperture corrections recommended
by the $Spitzer$ Science Center (see Table~4).

%%%%%%%%%%%%%%%%%%%%%%%%%%%%%%%%%%%%%%%%%%%%%%%%%%%%%%%%%%%%%%%%%%%%%%%%%%%%
\section{DISCUSSION}
%%%%%%%%%%%%%%%%%%%%%%%%%%%%%%%%%%%%%%%%%%%%%%%%%%%%%%%%%%%%%%%%%%%%%%%%%%%%

%%%%%%%%%%%%%%%%%%%%%%%%%%%%%%%%%%%%%%%%%%%%%%%%%%%%%%%%%%%%%%%%%%%%%%%%%%%%
\subsection{Spectral Energy Distributions and Evolutionary Stages}
%%%%%%%%%%%%%%%%%%%%%%%%%%%%%%%%%%%%%%%%%%%%%%%%%%%%%%%%%%%%%%%%%%%%%%%%%%%%

Figure~8 shows the spectral energy distributions (SEDs) of
SVS\,13\,A, B, and C. IRAS far-infrared
(only 100\,$\mu$m data are used here) data are adopted from
Jennings et al. (1987); SCUBA submm data from
Chandler \& Richer (2000); IRAM-30\,m 1.3\,mm data from Chini et
al. (1997)\footnote{We use the single-dish fluxes because the
interferometric 1.3\,mm (B98) and 1.4\,mm (this work)
data do not recover the envelope fluxes.};
BIMA 2.7\,mm data from LMW2000; IRAM PdBI 3.1\,mm data from
this work and 3.5\,mm data from B98. Since IRAS could not
resolve SVS\,13, the 100\,$\mu$m flux ratio of
A\,:\,B\,:\,C = 8:1:1 was inferred from the ratios at other
wavelengths. For source B, the
$Spitzer$ sensitivities at IRAC bands (also for source C) and
MIPS bands are adopted\footnote{see
http://ssc.spitzer.caltech.edu/irac/sens.html and
http://ssc.spitzer.caltech.edu/mips/sens.html}.

In order to derive luminosities and temperatures, we first
interpolated and then integrated the SEDs, always assuming
spherical symmetry. Interpolation between the flux densities was
done by a $\chi$$^2$ single-temperature grey-body fit to all
points at $\lambda$\,$\geq$\,100\,$\mu$m, using the method
described in Paper\,II. A simple logarithmic interpolation was
performed between all points at $\lambda$\,$\leq$\,100\,$\mu$m.
The fitting results are listed in Table~5.

The fitting results of sources B and C confirm the earlier
suggestion (see e.g., Chandler \& Richer 2000) that they
are Class\,0 protostars. For source A, the high bolometric
temperature ($\sim$ 114\,K) and low $L_{\rm submm}$/$L_{\rm bol}$
ratio ($\sim$ 0.8\%) suggest it is a Class\,I young stellar
object. Nevertheless, we want to note that source A resembles a
Class\,0 protostar in at least two aspects: (1) it is associated with
a cm radio source (VLA\,4; Rodr\'{i}guez et al. 1999) and embedded
in a large-scale dusty envelope, and (2) it is
driving an extremely high velocity CO outflow (B2000), which
is believed to be one of the characteristics of Class\,0
protostars (see Bachiller 1996). We speculate that source A
could be a Class\,0/I transition object, which is visible at
infrared wavelengths due to the high inclination angle (see B2000).

%%%%%%%%%%%%%%%%%%%%%%%%%%%%%%%%%%%%%%%%%%%%%%%%%%%%%%%%%%%%%%%%%%%%%%%%%%%%
\subsection{Gas Kinematics of the N$_{2}$H$^{+}$ Core}
%%%%%%%%%%%%%%%%%%%%%%%%%%%%%%%%%%%%%%%%%%%%%%%%%%%%%%%%%%%%%%%%%%%%%%%%%%%%

\subsubsection{Turbulence}

Assuming that the kinetic gas temperature is equal to the dust
temperature ($\sim$\,20\,K; see Table~5), the thermal line widths
of N$_{2}$H$^{+}$ and an ``average" particle of mass 2.33\,m$_{\rm
H}$ (assuming gas with 90\% H$_2$ and 10\% He) are
$\sim$\,0.18\,km\,s$^{-1}$ and $\sim$\,0.62\,km\,s$^{-1}$,
respectively. The latter line width represents the local sound
speed. The non-thermal contribution to the N$_{2}$H$^{+}$ line
width ($\triangle v_{\rm NT} = \sqrt{\triangle v_{\rm mean}^2 -
\triangle v_{\rm th}^2}$) is then estimated to be
$\sim$\,0.44\,km\,s$^{-1}$, which is about two times larger than
the thermal line width, but smaller than the local sound speed
(i.e., subsonic).

It is widely accepted that turbulence is the main contribution to
the non-thermal line width (Goodman et al. 1998), while infall,
outflow, and rotation could also broaden the line width in
protostellar cores. The spatial distribution of the line widths in
SVS\,13 (see Fig.\,5) shows that large line widths
($\geq$\,0.45\,km\,s$^{-1}$) only occur in the gap between sources
B and VLA\,3 and the jet-like extension at the edge of the core.
At or around the two sources, line widths are relatively small
(0.3\,$-$\,0.4\,km\,s$^{-1}$). Excluding the gap and extension,
the re-estimated non-thermal line width would be
$\sim$\,0.35\,km\,s$^{-1}$. Although this is still larger than
the thermal line width, it is fair to say that the turbulence in
the SVS\,13 core is at a low level.

\subsubsection{Core Rotation}

The velocity field map (Fig.\,4) shows a clear velocity gradient
across the core of $\sim$ 28\,km\,s$^{-1}$\,pc$^{-1}$ increasing from
southwest to northeast, i.e., roughly along the connection line
between sources B and VLA\,3. As discussed in Paper\,I, systematic
velocity gradients are usually dominated by either rotation or
outflow. As seen in Fig.\,4, the large angle ($>$\,70$^\circ$)
between the gradient and the outflow axis suggests that the
observed velocity gradient in N$_{2}$H$^{+}$ is due to core
rotation rather than outflow.

The velocity gradient derived in SVS\,13 is much larger than the
results found in Papers\,I\,\&\,II
($\sim$\,7\,km\,s$^{-1}$\,pc$^{-1}$) and other Class\,0 objects,
e.g., IRAM\,04191 ($\sim$\,17\,km\,s$^{-1}$\,pc$^{-1}$;
Belloche \& Andr\'{e} 2004), suggesting fast rotation of the
N$_{2}$H$^{+}$ core. With this gradient and the FWHM core size
($\sim$\,1520\,AU), the specific angular momentum $J/M$ is
calculated to be
$\sim$\,0.47\,$\times$\,10$^{-3}$\,km\,s$^{-1}$\,pc
($\sim$\,1.45\,$\times$\,10$^{16}$\,m$^2$\,s$^{-1}$), using the
same method as described in Paper\,II. The ratio of the rotational
energy to the gravitational potential energy is also calculated
with the same method as described in Paper\,II, and the estimated
$\beta_{\rm rot}$ value is $\sim$\,0.025 (uncorrected for inclination
angle).

%%%%%%%%%%%%%%%%%%%%%%%%%%%%%%%%%%%%%%%%%%%%%%%%%%%%%%%%%%%%%%%%%%%%%%%%%%%%
\subsection{SVS\,13 B and VLA\,3: A Bound Protobinary System}
%%%%%%%%%%%%%%%%%%%%%%%%%%%%%%%%%%%%%%%%%%%%%%%%%%%%%%%%%%%%%%%%%%%%%%%%%%%%

Our observations show a weak 3\,mm dust continuum source at the
position of the radio source VLA\,3 (see Fig.\,1). This dust
continuum source is embedded in the elongated N$_{2}$H$^{+}$
core together with source B (see Fig.\,2). The angular separation
and flux ratio between sources B and VLA\,3 are measured to be
10\farcs7\,$\pm$\,0\farcs2 and $\sim$\,13 from the 3\,mm dust
continuum image, respectively.

The velocity field map shows a clear velocity gradient across
the two sources, with sources VLA\,3 and B being located in the
red- and blue-shifted regions, respectively. As discussed above,
we assume that this systematic velocity gradient is due to the
core rotation. The two sources are thus probably forming a protobinary
system that has originated from rotational fragmentation of the core
that we now observe in the N$_{2}$H$^{+}$ line.

Fig.\,9 shows a position-velocity diagram, along the
connecting-line between sources B and VLA\,3. The radial
velocity difference between the two sources is
$\sim$\,0.5\,km\,s$^{-1}$. Assuming that the two sources are
in Keplerian rotation and that the orbit is circular
and perpendicular to the plane to the sky, the velocity difference
yields a combined binary mass of $\sim$\,1.1\,$M_\odot$. Assuming
a mass ratio of $\sim$\,13 ($M_{\rm B}$/$M_{\rm VLA3}$), the
derived masses of sources B and VLA3 are 1.0 and 0.08\,$M_\odot$,
respectively. The estimated dynamical mass of source B is consistent
with the total gas mass derived from the 1.4\,mm dust
continuum image (1.05\,$M_\odot$). This in turn supports our
assumption above that both sources B and VLA\,3 are forming a
bound binary system. Hereafter we refer to this binary protostar
as the SVS\,13\,B/VLA\,3 protobinary system. We also note that
source A, which is much closer to VLA\,3 in the 3\,mm dust
continuum image, is probably further away in foreground or
background. This shows that 2-D images can be very misleading
unless additional kinematic information like that derived from
our N$_2$H$^+$ observations is also considered.

%%%%%%%%%%%%%%%%%%%%%%%%%%%%%%%%%%%%%%%%%%%%%%%%%%%%%%%%%%%%%%%%%%%%%%%%%%%%
\subsection{Hierarchical Fragmentation in SVS\,13:\\
Initial Fragmentation vs. Rotational Fragmentation}
%%%%%%%%%%%%%%%%%%%%%%%%%%%%%%%%%%%%%%%%%%%%%%%%%%%%%%%%%%%%%%%%%%%%%%%%%%%%

The 3\,mm dust continuum image shows three distinct cores (A, B, and C)
in SVS\,13, which are roughly aligned along a line in the
northeast-southwest direction (see Fig.\,1).
The morphology of the N$_2$H$^+$ emission approximately follows this
alignment. Nevertheless, there is an emission gap (hole) between the
N$_2$H$^+$ main core and the small clumps towards northeast of source
A (see Fig.\,2). This gap is possibly caused by an effect of the outflow,
which releases great amounts of CO molecules and thus destroys N$_2$H$^+$
around source A (see Aikawa et al. 2001 and also discussion in Paper II).
The projected separations between sources A-B and B-C are $\sim$\,5000\,AU and
$\sim$\,7000\,AU, respectively. Previous single-dish
submm and mm maps show a large-scale common envelope with a radius
of $\sim$\,20000\,AU (0.1\,pc), surrounding these three cores and
elongated in the same direction (Chini et al. 1997;
Chandler \& Richer 2000).
Based on these associated morphologies,
we speculate that the three cores (A, B, and C) were formed by
initial fragmentation\footnote{The term `initial fragmentation'
refers to turbulent fragmentation in large-scale cores, prior to
the protostellar phase. This leads to the formation of individual
sub-cores, which are usually not gravitationally bound to each
other.} of a large-scale filamentary prestellar core.

On the other hand, in the SVS\,13\,B sub-core, a binary protostar
with a projected separation of 3800\,AU is
forming. The velocity field associated with this binary protostar
is dominated by rotation and the ratio of rotational energy to
gravitational energy is estimated to be
$\beta_{\rm rot}$ $\sim$\,0.025. The $\beta_{\rm rot}$ value
is believed to play an important role in the fragmentation process
(see reviews by Bodenheimer et al. 2000 and Tohline 2002).
Considering that magnetic fields support fragmentation,
Boss (1999) has shown that rotating cloud cores fragment when
$\beta_{\rm rot}$ $>$ 0.01 initially (but see also Machida et al.
2005). Based on this velocity structure, we suggest
that the binary protostar in SVS\,13 B is formed through
rotational fragmentation of a single collapsing protostellar
core.

Altogether we suggest a hierarchical fragmentation picture for
SVS\,13. (1) A large-scale filamentary prestellar core was
initially fragmented into three sub-cores (sources A, B, and C)
due to turbulence (see the review by Goodwin et al. 2007). (2)
These sub-cores are continually contracting toward higher degrees
of central condensation, and the rotation of the sub-cores is
getting faster and dominates the gas motion instead of turbulence.
(3) At a certain point, the condensations inside these sub-cores
become gravitationally unstable and start to collapse separately to
form either a binary protostar (e.g., SVS\,13\,A and B) if the rotational
energy is larger than a certain level (e.g., $\beta_{\rm rot}$ $>$
0.01) or a single protostar (e.g., SVS\,13\,C?) if the rotational
energy is not high enough to trigger the fragmentation.

%%%%%%%%%%%%%%%%%%%%%%%%%%%%%%%%%%%%%%%%%%%%%%%%%%%%%%%%%%%%%%%%%%%%%%%
\subsection{Accretion onto Protobinary: Simulations vs. Observations}
%%%%%%%%%%%%%%%%%%%%%%%%%%%%%%%%%%%%%%%%%%%%%%%%%%%%%%%%%%%%%%%%%%%%%%%

Accretion onto a protobinary is one of the most important
contributions to the final parameters of the system. Numerical
simulations of accretion have found that the main dependencies are
on the binary mass ratio and the relative specific angular
momentum, $j_{\rm inf}$, of the infalling material (see Bate \&
Bonnell 1997 and Bate 2000). For unequal mass components, low
$j_{\rm inf}$ material is accreted by the primary or its disk; for
higher $j_{\rm inf}$, the infalling material can also be accreted by the
secondary or its disk; when $j_{\rm inf}$ approaches the specific
angular momentum of the binary system $j_{\rm B}$, circumbinary disk
formation begins; when $j_{\rm inf}$ $\gg$ $j_{\rm B}$, the infalling
material is accreted by neither component but only forms a circumbinary
disk.

In our observations conducted at OVRO, ATCA and PdBI, we find in
several protobinary systems that only one component is associated
with a great amounts of circumstellar dust while the other is
somewhat naked, e.g., SVS\,13\,B/VLA\,3 (this work), CB\,230
(Launhardt 2001; Launhardt et al. in prep.), and BHR\,71
(Paper\,II). A similar situation was also found in the SVS\,13\,A
protobinary system (Anglada et al. 2004). In fact, there is a trend
derived from observations of only a very small number of sources
(i.e., not a statistically significant sample) that unequal
circumstellar masses (mass ratios below 0.5) protobinary systems are
much more common than those equal masses systems (Launhardt 2004; Chen
et al. in prep.). Since the detection bias goes towards equal-mass
protobinary systems (e.g., sources VLA\,1623, LMW2000 and L723 VLA2,
Launhardt 2004), we consider this trend is not negligible. The large
contrast between the two components implies that the accretion and
hence development of a circumstellar disk occurs preferentially in
only one component, with the disk absent or much less significant in
the other one.

According to the definition in Bate \& Bonnell (1997, hereafter BB97),
we can calculate the specific orbital angular momenta of binary system
as $j_{\rm B}$ = $\sqrt{GM_{\rm B}D}$ (where $G$, $M_{\rm B}$, and $D$
are the gravitational constant, total binary mass, and separation,
respectively), and consider it as an unity. Taking SVS\,13 B as an
example, we assume $M_{\rm B}$ = 1.0\,$M_\odot$ (see $\S$\,4.3) and
$D$ = 3800\,AU, and adopt the specifical angular momentum derived from
the N$_{2}$H$^{+}$ velocity field as $j_{\rm inf}$ ($\sim$
1.5\,$\times$\,10$^{16}$\,m$^2$\,s$^{-1}$). We then derive
$j_{\rm inf}$/$j_{\rm B}$ $\sim$\,0.1. This ratio is exactly
corresponding to the simulated case that only circumprimary disk
could be formed in an accreting protobinary system (see BB97).
Therefore, the fact that only one component in a protobinary system
is associated with a large amount of circumstellar material (or a
circumstellar disk) could be explained by the relatively low specific
angular momentum in the infalling envelope.

For a comparison, we list in Table~6 the corresponding parameters of
L1551 IRS5, a well-studied protobinary system which we already knew
the information of infalling gas and binary parameters (e.g.,
dynamical mass and separation). In contrast to SVS\,13B, L1551 IRS5
apparently corresponds to a case of accretion of material with relatively
higher angular momentum ($j_{\rm inf}$/$j_{\rm B}$\,$\sim$\,1.0), and
both components in the system are observed with a circumstellar disk
(see Rodr\'{i}guez et al. 1998), which is also consistent with the
theoretical simulation in BB97. Nevertheless, so far the data which would
reveal the link between specific angular momenta and formation of
circumstellar disks are very rare, and no statistically significant
correlations can be derived yet.

%%%%%%%%%%%%%%%%%%%%%%%%%%%%%%%%%%%%%%%%%%%%%%%%%%%%%%%%%%%%%%%%%%%%%%%
\section{CONCLUSION}
%%%%%%%%%%%%%%%%%%%%%%%%%%%%%%%%%%%%%%%%%%%%%%%%%%%%%%%%%%%%%%%%%%%%%%%

We present IRAM-PdBI observations in the N$_{2}$H$^{+}$(1\,$-$\,0)
line and the 1.4 and 3\,mm dust continuum toward the low-mass
protostellar core SVS\,13. Complementary infrared data from the
{\it Spitzer Space Telescope} are also used in this work. The main
results are summarized below:

(1) The 3\,mm dust continuum image resolves four sources in
SVS\,13, named A, B, C, and VLA\,3, respectively. Source
C lies outside of the view at 1.4\,mm, while source VLA\,3 is not
seen in the 1.4\,mm continuum image. The separations between
sources A-B and B-C are measured to be $\sim$\,5000\,AU and
$\sim$\,7000\,AU, respectively. Assuming optically thin dust
emission, we derive total gas masses of
$\sim$\,0.2\,$-$\,1.1\,$M_\odot$ for the sources.

(2) N$_{2}$H$^{+}$\,(1--0) emission is detected in SVS\,13.
The emission is spatially associated with the dust continuum
sources B and VLA\,3. The excitation temperature of the
N$_{2}$H$^{+}$ line is $\sim$\,11\,K. The FWHM radius of the
N$_{2}$H$^{+}$ core is $\sim$\,1520\,AU. The line widths are
roughly constant within the interiors of the core and large line
widths only occur in the gap between the two dust continuum
sources and a jet-like extension at the edge of the core.
The observed mean line width is $\sim$\,0.48\,km\,s$^{-1}$ and
the derived virial mass of the N$_{2}$H$^{+}$ core is
$\sim$\,0.7\,$M_\odot$.

(3) We derive the N$_{2}$H$^{+}$ radial velocity field for
SVS\,13. The velocity field shows a systematic velocity gradient
of $\sim$\,28\,km\,s$^{-1}$\,pc$^{-1}$ across the dust continuum
sources B and VLA\,3, which could be explained by rotation. We
estimate the specific angular momentum of
$\sim$\,0.47\,$\times$\,10$^{-3}$\,km\,s$^{-1}$\,pc
($\sim$\,1.45\,$\times$\,10$^{16}$\,m$^{2}$\,s$^{-1}$) for this
N$_{2}$H$^{+}$ core. The ratio of rotational energy to
gravitational energy is $\sim$\,0.025.

(4) Infrared emission from sources A and C is detected at
$Spitzer$ IRAC bands and MIPS bands. Source A is driving a chain
of HH objects, as seen in the IRAC 4.5\,$\mu$m infrared
image. No infrared emission is detected from source B in
any $Spitzer$ bands, suggesting the source is deeply embedded.

(5) By fitting the spectral energy distributions, we derive the
dust temperature, bolometric temperature, and bolometric
luminosity for the sources A, B, and C. We find that sources
B and C are Class\,0 protostars, while source A could be a
Class\,0/I transition object.

(6) The velocity field associated with sources B and VLA\,3
suggests that the two sources are forming a physically bound
protobinary system embedded in a common N$_{2}$H$^{+}$ core.
The estimated dynamical mass of source B
($\sim$\,1.0\,$M_\odot$) is consistent with its gas mass
derived from the mm dust continuum images.

(7) Based on the morphologies and velocity structures, we suggest
a hierarchical fragmentation picture for SVS\,13: the three
sources (A, B, and C) in SVS\,13 were formed by initial
fragmentation of a filamentary prestellar core, while the
protobinary systems SVS\,13\,B/VLA\,3 and SVS\,13\,A were
formed by rotational fragmentation of the collapsing sub-cores.

(8) Our observations conducted at OVRO, ATCA, and PdBI find in
several protobinary systems that only one component is associated
with a large amount of circumstellar material while the other is
somewhat naked, implying that the accretion and hence development
of a circumstellar disk occurs preferentially in only one
component, with the disk absent or much less significant in the
other one. We find that this situation could be explained by the
relatively low specific angular momentum in the infalling
envelope, which is supported by both simulations and observations.

\begin{acknowledgements}

We thank the anonymous referee for many insightful comments and
suggestions. We want to thank the IRAM staff for help provided
during the observations and data reduction. X. Chen thanks the
research funding from the European Community's Sixth Framework
Programme for data reduction travel at the IRAM/Grenoble.
\end{acknowledgements}

\clearpage
%%%%%%%%%%%%%%%%%%%%%%%%%%%%%%%%%%%%%%%%%%%%%%%%%%%%%%%%%%%%%
%\bibliographystyle{apj}                        %% AASTeX
%\bibliography{protobinary.bbl}                 %% includes the journal abbrevs

\clearpage
%%%%%%%%%%%%%%%%%%%%%%%%%%%%%%%% Table %%%%%%%%%%%%%%%%%%%%%%%%%%%%%%%%%%%%%%%%%%
%%%%%%%%%%%%%%%%%%%%%%%%%% Table 1: PdBI Observational Log %%%%%%%%%%%%%%%%%%%%%%
\begin{deluxetable}{ccccccc}
\tabletypesize{\scriptsize} \tablecaption{\footnotesize IRAM-PdBI
observation logs of SVS13.\label{pdbi}} \tablewidth{0pt}
\tablehead{\colhead{Object} &\colhead{Other}&\colhead{R.A. \& Dec.
(J2000)$^{a}$} &\colhead{Distance}&\colhead{Array}
&\colhead{HPBW$^{b}$} &\colhead{rms$^c$}\\
\colhead{Name}&\colhead{Name}&\colhead{[h\,:\,m\,:\,s,
$^{\circ}:\,':\,''$]}&\colhead{[pc]}&\colhead{configuration}
&\colhead{[arcsecs]} &\colhead{[mJy/beam]}}\startdata
SVS13 & HH\,7$-$11 & 03:29:03.20, 31:15:56.00  &350& CD   & 2.8$\times$2.5, 1.2$\times$1.1 & 15.5, 0.62, 3.7\\
\enddata
\tablenotetext{a}{Reference position for figures and tables in the
paper.} \tablenotetext{b}{Naturally-weighted synthesized FWHM beam
sizes at 3\,mm and 1.4\,mm dust
continuum.}\tablenotetext{c}{1\,$\sigma$ noises at
N$_2$H$^+$(1--0) line, 3\,mm dust continuum, and 1.4\,mm dust
continuum.}
\end{deluxetable}
%%%%%%%%%%%%%%%%%%%%%%%%%%%%%%%%%%%%%%%%%%%%%%%%%%%%%%%%%%%%%%%%%%%%%%%%%%%%%%%%

%%%%%%%%%%%%%%%%%%%%%%%%%%%%% Table 2: PdBI mmc results %%%%%%%%%%%%%%%%%%%%%%%%%%%
\begin{deluxetable}{rllccccccc}
\tabletypesize{\scriptsize} \tablecaption{\footnotesize PdBI mm
dust continuum results for SVS\,13.\label{mmc}} \tablewidth{0pt}
\tablehead{\colhead{Source}&\multicolumn{2}{c}{Position$^{a}$}&
\colhead{}&\multicolumn{2}{c}{$S_{\nu}$$^{a,b}$ [mJy]} & \colhead{}&
\multicolumn{2}{c}{FWHM sizes at 3\,mm$^a$} & \colhead{$M_{\rm gas}$}\\
\cline{2-3}\cline{5-6}\cline{8-9}\colhead{}&
\colhead{$\alpha$(J2000)}& \colhead{$\delta$(J2000)} & \colhead{}&
\colhead{$\lambda$3\,mm} & \colhead{$\lambda$1.4\,mm}& \colhead{}
& \colhead{maj.$\times$min.}& \colhead{P.A.} &
\colhead{[$M_{\odot}$]} } \startdata
SVS\,13\,A     & 03:29:03.75 & 31:16:03.76 && 30.0$\pm$4.5 &  135$\pm$21  && 2\farcs4$\times$1\farcs4 &--58$\pm$6\degr   & 0.75$\pm$0.12   \\
B              & 03:29:03.07 & 31:15:52.02 && 42.4$\pm$6.4 &  193$\pm$30  && 2\farcs7$\times$1\farcs8 &--46$\pm$5\degr   & 1.05$\pm$0.16   \\
C              & 03:29:01.96 & 31:15:38.26 &&  6.5$\pm$1.0 &  $--$        && $\sim$\,0\farcs7         & $--$             & 0.20$\pm$0.03   \\
\enddata
\tablenotetext{a}{Derived from
Gaussian $uv$ plane fitting.} \tablenotetext{b}{The error bar is
derived from $\sqrt{\sigma^2_{\rm cali}+\sigma^2_{\rm fit}}$,
where $\sigma_{\rm cali}$ is the uncertainty from calibration
($\sim$\,15\% of flux density) and $\sigma_{\rm fit}$ is the
uncertainty from Gaussian fitting.}
\end{deluxetable}
%%%%%%%%%%%%%%%%%%%%%%%%%%%%%%%%%%%%%%%%%%%%%%%%%%%%%%%%%%%%%%%%%%%%%%%%%%%%%%%%%%%

%%%%%%%%%%%%%%%%% Observing Parameters from the HFS Fitting %%%%%%%%%%%%%%%%%%%%%%%

\begin{deluxetable}{ccccc}
\tabletypesize{\scriptsize}\tablecaption{\footnotesize Results of
N$_2$H$^+$\,(1--0) spectra fitting$^{a}$\label{n2hp}} \tablewidth{0pt}
\tablehead{\colhead{Peak [$''$]} & \colhead{$V_{\rm LSR}$ [km
s$^{-1}$]} & \colhead{$\triangle$$v$
[km\,s$^{-1}$]}&\colhead{$\tau_{\rm tot}$} & \colhead{$T_{\rm ex}$
[K]} } \startdata
(2, 4)        & 8.90$\pm$0.01   & 0.31$\pm$0.01   &  1.9$\pm$0.1   & 11.3$\pm$0.5 \\
($-$2, $-$4)  & 8.39$\pm$0.01   & 0.46$\pm$0.02   &  2.2$\pm$0.1   & 10.1$\pm$0.5 \\
\enddata
\tablenotetext{a}{Value at the two intensity peaks. The error
represents 1 $\sigma$ error in the hyperfine fitting.}
\end{deluxetable}
%%%%%%%%%%%%%%%%%%%%%%%%%%%%%%%%%%%%%%%%%%%%%%%%%%%%%%%%%%%%%%%%%%%%%%%%%%%%%%%%%%%%%

%%%%%%%%%%%%%%%%%%%%%%%%%%%%%%%%%%%%%%%%%%%%%%%%%%%%%%%%%%%%%%%%%%%%%%%%%%%%%%%%%%%%%
\begin{deluxetable}{rllcccccc}
\tabletypesize{\scriptsize} \tablecaption{\footnotesize $Spitzer$
flux densities of SVS\,13 A and SVS\,13 C. \label{spitzer}}
\tablewidth{0pt} \tablehead{\colhead{} &\colhead{R.A.$^a$}
&\colhead{Dec.$^a$} &\colhead{$S(3.6\,\mu m)$}
&\colhead{$S(4.5\,\mu m)$} &\colhead{$S(5.8\,\mu m)$}
&\colhead{$S(8.0\,\mu m)$} &\colhead{$S(24\,\mu m)$}&\colhead{$S(70\,\mu m)$}\\
\colhead{Source} &\colhead{(J2000)} &\colhead{(J2000)}
&\colhead{[mJy]} &\colhead{[mJy]} &\colhead{[mJy]}
&\colhead{[mJy]} &\colhead{[Jy]}&\colhead{[Jy]}}\startdata
SVS\,13 A          & 03:29:03.73 & 31:16:03.80 & 292$\pm$20  & 499$\pm$44  & 2470$\pm$140   & 2330$\pm$100   & $>$\,21       & 112$\pm$3\\
        C          & 03:29:01.98 & 31:15:38.17 & $<$\,0.36$^b$  & $<$\,0.43$^b$  & $<$\,1.3$^b$  & $<$\,0.54$^b$ & 0.22$\pm$0.01 &  $>$\,12\\
\enddata
\tablenotetext{a}{Peak positions measured at IRAC band 3 (for
SVS\,13 A) and MIPS band 1 (for SVS\,13
C).}\tablenotetext{b}{$Spitzer$ IRAC sensitivities.}
\end{deluxetable}
%%%%%%%%%%%%%%%%%%%%%%%%%%%%%%%%%%%%%%%%%%%%%%%%%%%%%%%%%%%%%%%%%%%%%%%%%%%%%%%%%%%%%

%%%%%%%%%%%%%%%%%%%%%%%%%%% Classification %%%%%%%%%%%%%%%%%%%%%%%%%%%%%%%%%%%%%%%%%%
\begin{deluxetable}{rcccccc}
\tabletypesize{\scriptsize} \tablecaption{\footnotesize Fitting
results of the spectral energy distribution} \label{sed}
\tablewidth{0pt} \tablehead{\colhead{Source}&\colhead{$T_{\rm
dust}$}&\colhead{$T_{\rm bol}$}&\colhead{$L_{\rm
bol}$}&\colhead{$L_{\rm submm}$}
&\colhead{$L_{\rm submm}$/$L_{\rm bol}$}&\colhead{Class}\\
\colhead{}&\colhead{[K]}&\colhead{[K]}&\colhead{[$L_\odot$]}&\colhead{[$L_\odot$]}&\colhead{[\%]}&\colhead{}}\startdata
SVS\,13 A        & 33 & 114    & 46.6 & 0.37    & 0.8 & 0/I(?)\\
        B        & 22 & 28     &  5.6 & 0.28    & 5.0 & 0 \\
        C        & 20 & 36     &  4.9 & 0.13    & 2.7 & 0 \\
\enddata
\end{deluxetable}
%%%%%%%%%%%%%%%%%%%%%%%%%%%%%%%%%%%%%%%%%%%%%%%%%%%%%%%%%%%%%%%%%%%%%%%%%%%%%%%%%%%%

\begin{deluxetable}{lcclrccc}
\tabletypesize{\scriptsize}\tablecaption{\footnotesize Specific
angular momenta of envelopes and binaries} \tablewidth{0pt}

\tablehead{\colhead{Source}&\colhead{Class}&\colhead{$j_{\rm inf}$$^a$}&\colhead{Mass$^b$}
&\colhead{Sepa.}&\colhead{$j_{\rm B}$} &\colhead{$j_{\rm inf}$/$j_{\rm B}$}&\colhead{Refs.$^c$}\\
\colhead{}&\colhead{}  &\colhead{[m$^{2}$ s$^{-1}$]} &\colhead{[$M_\odot$]}&\colhead{[AU]} &\colhead{[m$^{2}$ s$^{-1}$]} &\colhead{} &\colhead{}} \startdata
%\hline
L1551\,IRS5       &I &3.1$\times$$10^{16}$ & 1.2 (0.4)     & 40   &3.0$\times$$10^{16}$ & $\sim$1.0   &1,2\\
SVS\,13\,B        &0 &1.5$\times$$10^{16}$ & 1.1 ($<$0.1)  & 3800 &2.7$\times$$10^{17}$ & $\sim$0.1   &3\\
\enddata
\tablenotetext{a}{Specific angular momenta of infalling envelopes. For source L1551\,IRS5,
the rotation radius and velocity are estimated to be 900\,AU and 0.23\,km\,s$^{-1}$,
respectively (see Saito et al. 1996).}
\tablenotetext{b}{Total mass of the binary systems. The mass of L1551\,IRS5 is dynamical
mass estimated from orbital motion (see Rodr\'{i}guez 2004).}
\tablenotetext{c}{References--(1) Saito et al. 1996; (2) Rodr\'{i}guez 2004; (3) This work.}
\end{deluxetable}

%%%%%%%%%%%%%%%%%%%%%%%%%%%%%%%%%%%%%%%%%%%%%%%%%%%%%%%%%%%%%%%%%%%%%%%%%%%%%%%%%%%%
\clearpage
%%%%%%%%%%%%%%%%%%%%%%%%%%%% Figures %%%%%%%%%%%%%%%%%%%%%%%%%%%%%%%%%%%%%%%%%%%%%%%

\begin{figure*}
\begin{center}
\includegraphics[width=16cm,angle=0]{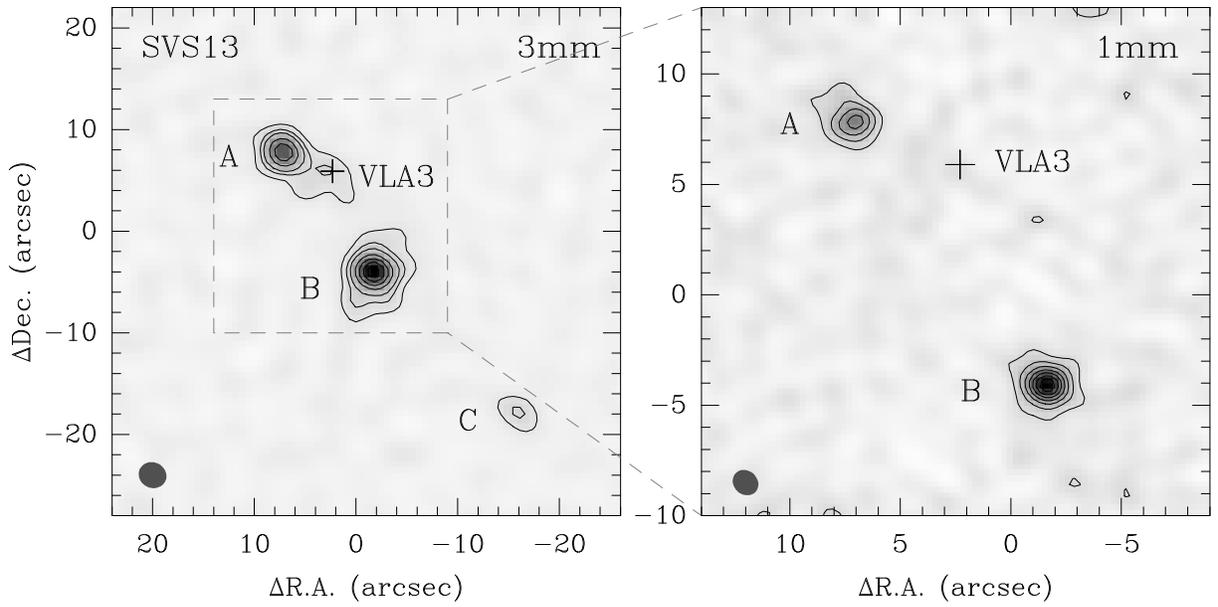}
\caption{IRAM-PdBI mm dust continuum images of SVS\,13. Contours
start at $\sim$\,3\,$\sigma$ (see Table~1) by steps of
$\sim$\,2\,$\sigma$. Crosses in the images mark the position of
the radio source VLA\,3 (Rodr\'{i}guez et al. 1997). The
synthesized PdBI beam is shown as grey oval in the
images.\label{fig1_mmc}}
\end{center}
\end{figure*}

\begin{figure*}
\begin{center}
\includegraphics[width=11cm,angle=0]{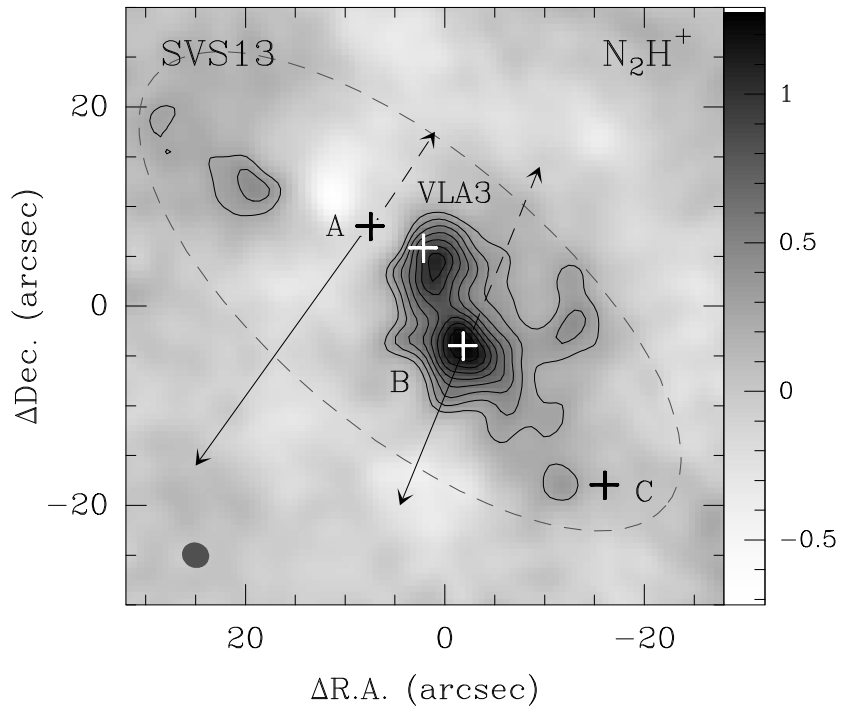}
\caption{Image of the N$_2$H$^+$(1--0) intensity integrated
over the seven hyperfine line components for SVS\,13. The unit of
the scale is [Jy beam$^{-1}$ km s$^{-1}$]. Contours start at
$\sim$\,3\,$\sigma$ (see Table~1) with steps of
$\sim$\,2\,$\sigma$. Crosses represent the peaks of 3\,mm dust
continuum emission. Solid and dashed arrows show the direction of
blue- and red-shifted outflows (see B98 and B2000). Dashed elliptical
contour shows an approximate size of the large-scale common envelope.
The synthesized PdBI beam is shown as grey oval.\label{fig2_n2hp}}
\end{center}
\end{figure*}

\begin{figure*}
\begin{center}
\includegraphics[width=10cm, angle=-90]{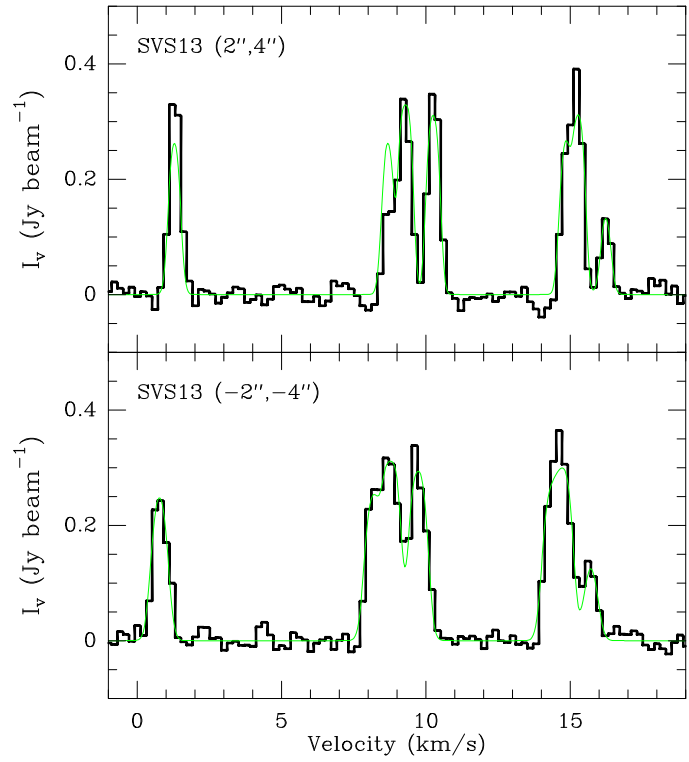}
\caption{N$_2$H$^+$ spectra at the two peak positions. Dotted
curves show the results of hyperfine structure line fitting (see a
color figure in the electronic version). Fit parameters are
given in Table~3.\label{fig3_spectra}}
\end{center}
\end{figure*}

\begin{figure*}
\begin{center}
\includegraphics[width=11cm]{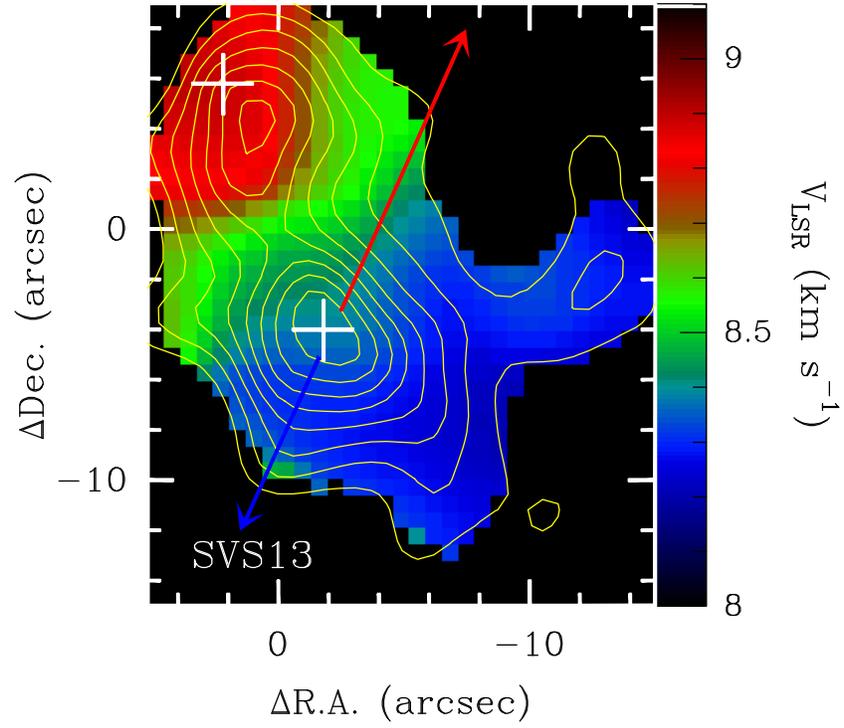}
\caption{N$_2$H$^+$ velocity field in SVS\,13. Contours and
crosses are the same as in Fig.\,2. Red and blue arrows show the
directions of red- and blue-shifted CO
outflows.\label{fig4_velocity}}
\end{center}
\end{figure*}

\begin{figure*}
\begin{center}
\includegraphics[width=12cm]{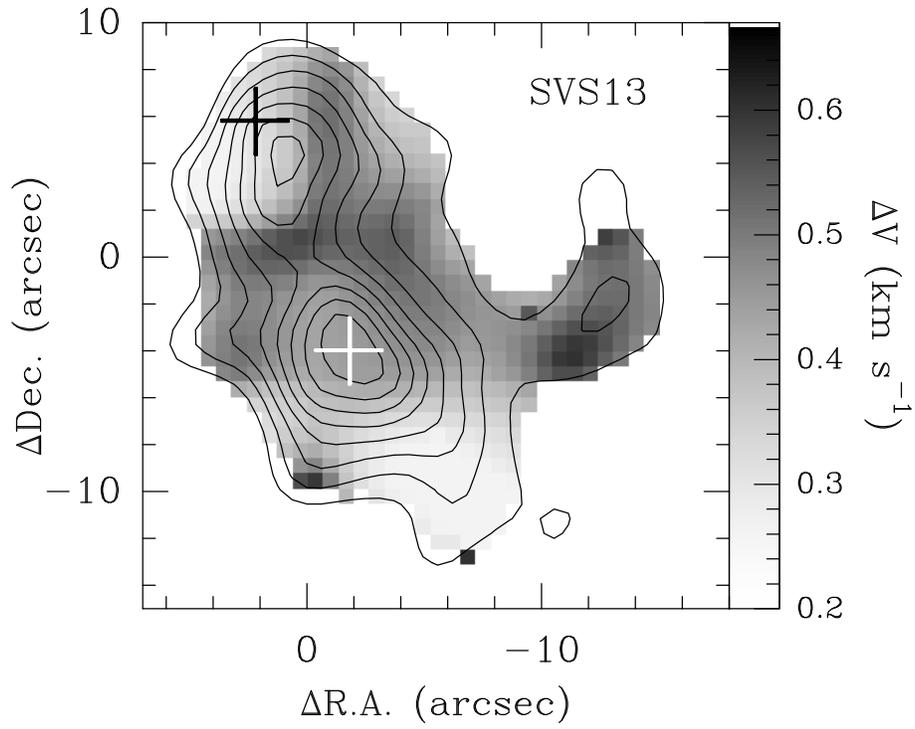}
\caption{Spatial distribution of N$_2$H$^+$ line widths in
SVS\,13, as derived from the HFS line fitting. Contours and
crosses are the same as them in Fig.\,2.\label{fig5_linewidth1}}
\end{center}
\end{figure*}

\begin{figure*}
\begin{center}
\includegraphics[height=10cm]{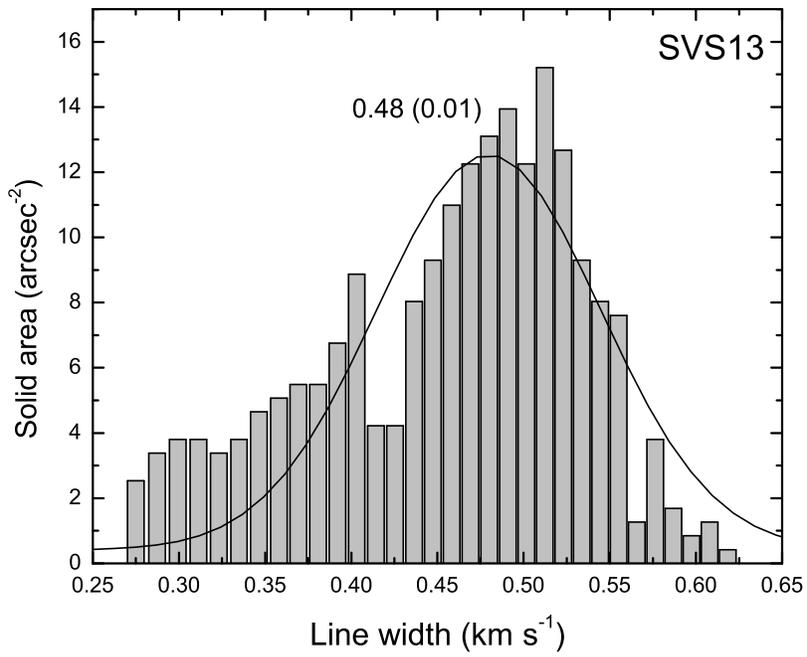}
\caption{Distribution of N$_2$H$^+$ line widths versus solid angle
areas for SVS\,13. Black solid curve and number show the results
of Gaussian fitting to the distribution.\label{fig6_linewidth2}}
\end{center}
\end{figure*}

\begin{figure*}
\begin{center}
\includegraphics[width=16cm,angle=0]{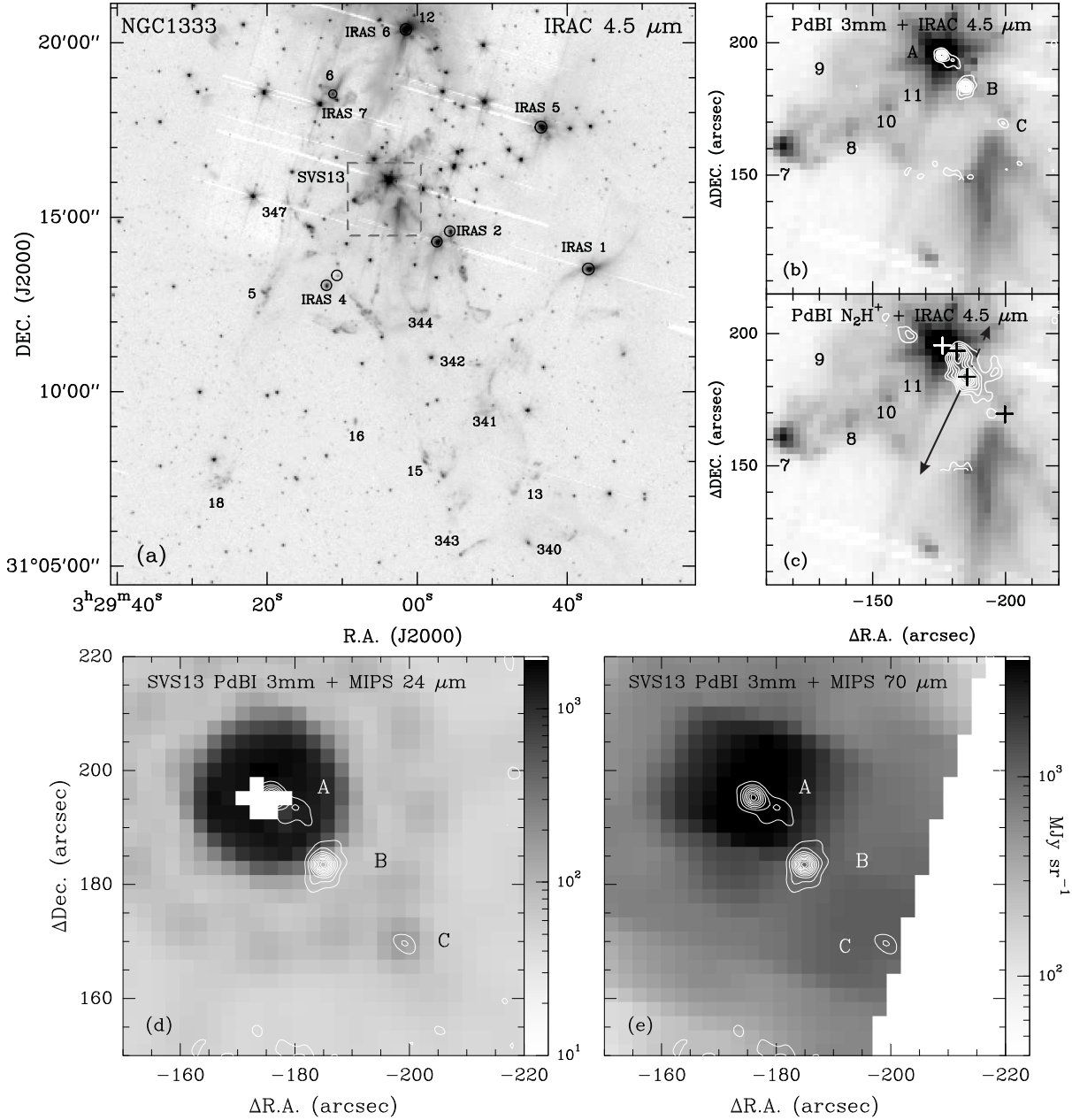}
\caption{$Spitzer$ images of SVS\,13. (a) $Spitzer$ IRAC band\,2
(4.5\,$\mu$m) image of SVS\,13 (reference position at
R.A.=03:29:17.49, DEC=31:12:48.54, J2000); (b) IRAC band\,2 image
overlaid with the PdBI 3\,mm dust continuum contours; (c) Same as
Fig.\,7b, but overlaid with the PdBI N$_2$H$^+$ intensity
contours; (d) $Spitzer$ MIPS\,1 (24\,$\mu$m) image of SVS\,13,
overlaid with the PdBI 3\,mm dust continuum contours; (e) Same as
Fig.\,7d, but for $Spitzer$ MIPS\,2 (70\,$\mu$m)
image.\label{fig7_spitzer}}
\end{center}
\end{figure*}

%\begin{figure*}
\begin{sidewaysfigure}
\begin{center}
\includegraphics[width=20cm, angle=0]{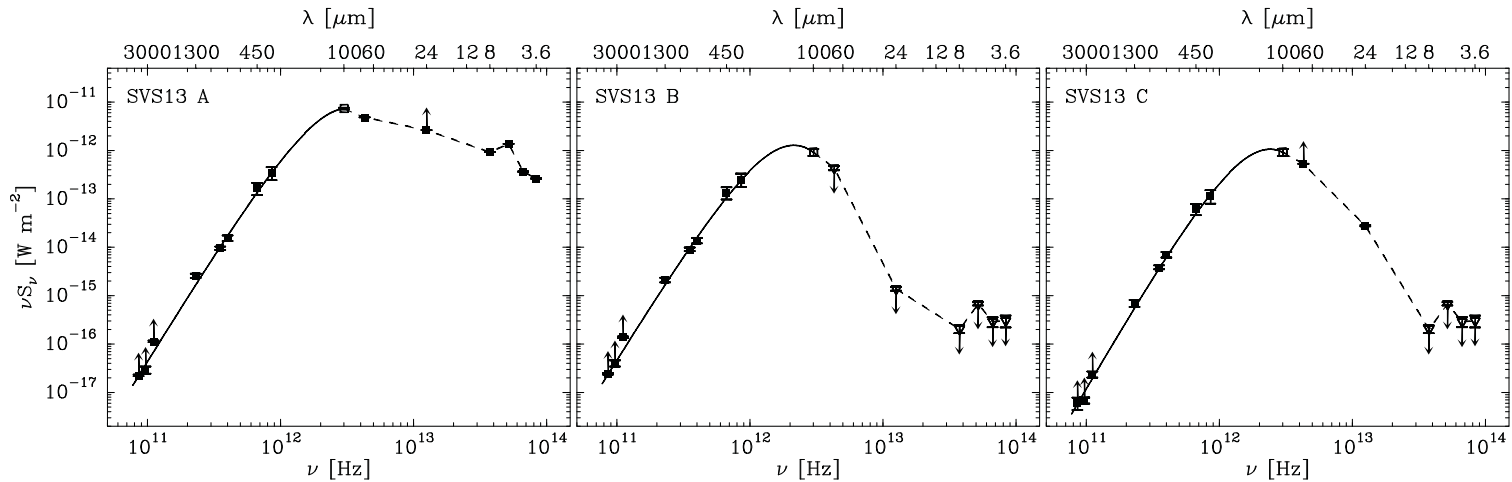}
\caption{Spectral energy distributions of SVS\,13\,A (left), B
(middle), and C (right). Error bars (1\,$\sigma$) are indicated
for all data points, but are mostly smaller than the symbol sizes.
While most solid squares represent real observational data points,
the 2.7$-$3.5\,mm fluxes were measured from interferometric maps
which resolved out the envelope and thus represent lower limits only.
Open squares represent IRAS 100\,$\mu$m data points, where flux
densities are estimated for the three sources with ratios assumed
in $\S$\,4.1. Open triangles represent the sensitivities of
IRAC and MIPS. Solid lines show the best-fit for all points at
$\lambda$ $\geq$ 100\,$\mu$m using a grey-body model (see text).
Dashed lines at $\lambda$\,$\leq$\,100\,$\mu$m show the simple
logarithmic interpolation used to derive the luminosity. The
fitting results are summarized in Table~5. \label{fig8_sed}}
\end{center}
\end{sidewaysfigure}
%\end{figure*}

\begin{figure*}[htcp]
\begin{center}
\includegraphics[width=10cm, angle=0]{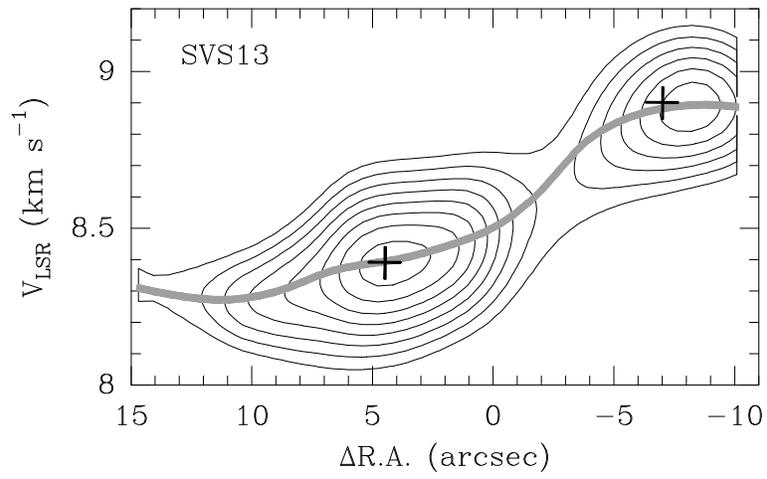}
\caption{Position-velocity diagram of SVS\,13 (along the
connecting line between sources B and VLA\,3). Crosses show the
locations of sources B and VLA\,3. The grey line marks the mean radial
velocity.\label{fig9_pv}}
\end{center}
\end{figure*}

\end{document}